\documentclass[showpacs,preprintnumbers]{revtex4}
\usepackage{amsmath,amssymb}
\usepackage{graphicx}
\usepackage{dcolumn}
\usepackage[vcentermath]{youngtab}
\usepackage[dvips]{epsfig,color}
\usepackage{bm}
\begin{document}
\title{Interpretation of the $X(3872)$ as a charmonium state plus an extra component due to the coupling to the meson-meson continuum}
\author{J. Ferretti}\author{G. Galat\'a}
\affiliation{INFN, Sezione di Genova, via Dodecaneso 33, 16146 Genova (Italy)}
\affiliation{Universidad Nacional Aut\'onoma de M\'exico, 04510 M\'exico DF, M\'exico}
\author{E. Santopinto}\thanks{Corresponding author: santopinto@ge.infn.it}
\affiliation{INFN, Sezione di Genova, via Dodecaneso 33, 16146 Genova (Italy)}    
\begin{abstract}
We present a quark model calculation of the charmonium spectrum with self energy corrections due to the coupling to the meson-meson continuum. 
The bare masses used in the calculation are computed within the relativized quark model by Godfrey and Isgur. 
The strong decay widths of $3S$, $2P$, $1D$ and $2D$ $c \bar c$ states are also calculated, in order to set the values of the $^3P_0$ pair-creation model's parameters we use to compute the vertex functions of the loop integrals. 
Finally, the nature of the $X(3872)$ resonance is analyzed and the main possibilities ($c \bar c$ state or $D \bar D^*$ molecule) are discussed. 
According to our results, the $X(3872)$ is compatible with the meson $\chi_{c1}(2P)$, with $J^{PC} = 1^{++}$, and is thus interpreted as a $c \bar c$ core plus higher Fock components due to the coupling to the meson-meson continuum. These $J^{PC} = 1^{++}$ quantum numbers are in agreement with the experimental results found by the LHCb collaboration. 
In our view, the $X(3872)$'s mass is lower than the quark model's predictions because of self energy shifts.
\end{abstract}
\maketitle

\section{Introduction}
The quark model (QM), in all its possible reformulations \cite{Isgur:1979be,Capstick:1986bm,Giannini:2001kb,Glozman-Riska,Bijker:1994yr,Loring:2001kx,Godfrey:1985xj,Iachello:1991re,Ferretti:2011,Santopinto:2006my,Valcarce:2005rr}, can properly describe several properties of the hadrons, such as the spectrum and the magnetic moments, but it neglects continuum coupling effects.
Indeed since the earliest days of hadron spectroscopy, it has been recognized that properties of levels can be strongly influenced by nearby channels \cite{Patrignani:2012an}.
The presence of these higher Fock components in meson and baryon wave functions are predicted by the QCD and must have an effect on the QM similar to that of unquenching lattice QCD calculations.
In particular, these continuum coupling effects can contribute, through a self-energy term, to a shift in the hadron masses, as already shown by several authors in the baryon \cite{Tornqvist,Blask:1987yv,Brack:1987dg,SilvestreBrac:1991pw,Horacsek:1986fz,Fujiwara:1992yv,Morel:2002vk} and meson \cite{Danilkin:2009hr,Danilkin:2010cc,Kalashnikova:2005ui,Hanhart:2007yq,Baru:2011rs,Ono:1983rd,Barnes:2007xu,Hwang:2004cd,Eichten:2004uh,Rupp:2006sb,Pennington:2007xr,bottomonium} sectors.

Interest in loop corrections in the meson sector \cite{Hwang:2004cd} was triggered after the discovery of the narrow charmed-strange mesons $D_{s0}^*(2317)^+$ \cite{Aubert:2003fg} and $D_{s1}(2460)^+$ \cite{Besson:2003cp}, since their surprisingly low masses could be explained by this type of effect.
In the 80's, T\"ornqvist $et$ $al.$ \cite{Ono:1983rd} studied heavy $c \bar c$ and $b \bar b$ quarkonium within the unitarized quark model and calculated the mass shifts and mixing induced by $D \bar D$, $D^* \bar D^*$, ... loop diagrams, using the $^3P_0$ decay model \cite{3P0} for hadron vertex functions.

Barnes and Swanson \cite{Barnes:2007xu} computed the mass shifts of charmonium $1S$, $2S$ and $1P$ resonances due to $D \bar D$, $D \bar D^*$, $D^* \bar D^*$, $D_s \bar D_s$, $D_s \bar D_s^*$, $D_s^* \bar D_s^*$ loops. The authors evaluated the coupling between the valence component and the continuum component by using the $^3P_0$ model \cite{3P0}, with Gaussian meson wave functions.

Danilkin and Simonov analyzed the mass shifts of charmonium $N^3S_1$ ($N = 1, 2, 3$) \cite{Danilkin:2009hr} and $2^{3,1}P_J$ \cite{Danilkin:2010cc} states, using the mechanism of channel coupling via decay products. 
The authors applied the Weinberg eigenvalue method \cite{Weinberg:1963zz} to multichannel problems, considering $D \bar D$, $D \bar D^*$ and $D^* \bar D^*$ decay channels. 

Eichten $et$ $al.$ \cite{Eichten:2004uh} evaluated the influence of open-charm channels on charmonium properties, such as strong decay widths and self energies. The authors revisited the properties of charmonium levels, using the Cornell coupled-channel model \cite{Eichten:1978tg} to assess departures from the single channel potential-model expectations.

Hwang and Kim \cite{Hwang:2004cd} calculated the mass shift of $D_{sJ}^*(2317)$ due to coupled channel effects, within the Cornell coupled-channel model of Ref. \cite{Eichten:1978tg}. 
According to them, the measured mass of this meson, being 160 MeV lower than the corresponding estimation of Ref. \cite{Godfrey:1985xj}, appears surprisingly low and can only be explained by coupled channel effects.
 
The loop corrections can be relevant to the study of the $X(3872)$ meson \cite{Choi:2003ue}, whose nature has not yet been understood. 
Indeed, there are currently two possible interpretations for the meson: a weakly-bound $1^{++}$ $D \bar D^*$ molecule \cite{Danilkin:2010cc,Hanhart:2007yq,Baru:2011rs,Swanson:2003tb,Fulsom:2007ve} or a $c \bar c$ state \cite{Meng:2007cx,Suzuki:2005ha}, with $1^{++}$ or $2^{-+}$ quantum numbers. 
For a summary of theoretical interpretations of the $X(3872)$, see Ref. \cite{Swanson:2006st}.

In the last few years, interest in heavy meson physics has increased enormously, as has the number of collaborations devoted to the topic. 
In particular, BaBar \cite{Prencipe:2012kb,Santoro:2012jq},  Belle \cite{Lange:2011qd}, CDF \cite{Yi:2009pz} and D0  have already provided many interesting results; moreover, all four detectors at LHC (Alice, Atlas, CMS and LHCb) have the capacity to study charmonia and bottomonia and have already produced some results, such as the discovery of a $\chi_b(3P)$ system \cite{Aad:2011ih}. There are also approved proposals for new experiments, such as Belle II \cite{Yuan:2012jd}.

The calculation presented in this article is the first attempt to calculate in a systematic way the spectrum of charmonia within a quark model, including loop corrections, and makes it possible to perform a comparison with the already existing and the future experimental data. 
Something similar has already been done for bottomonia in Refs. \cite{bottomonium}. 

Our results for the spectrum of charmonia are fitted to the experimental data \cite{Nakamura:2010zzi}, so that the calculated masses of the mesons of interest are the sum of a bare energy term, computed within the relativized QM by Godfrey and Isgur \cite{Godfrey:1985xj}, and a self energy correction, computed thanks to the formalism of the  unquenched quark model (UCQM) \cite{bottomonium,Santopinto:2010zza}.
In our UCQM calculation, we consider as intermediate states a complete set of accessible SU$_{\mbox{f}}$(4) ground-state (i.e. $1S$) mesons. 
$1S$ intermediate states, being at lower energies than $P$-wave and $D$-wave intermediate meson states, give the main contribution to the self energies of the charmonium states that we are going to study. 

Furthermore, we present some results for the strong decay widths of charmonium $3S$, $2P$, $1D$ and $2D$ states, calculated within a modified version of the $^3P_0$ pair-creation model \cite{3P0}. 
This is done in order to set the values of the $^3P_0$ model's parameters we use to compute the vertex function of the UCQM [see  Eq. (\ref{eqn:self-a})].

Finally, we use our results for the $c \bar c$ spectrum to discuss the nature of the $X(3872)$ resonance. 
Specifically, we analyze the interpretation of this meson as a $c \bar c$ state with $1^{++}$ or $2^{-+}$ quantum numbers. According to our results, the $X(3872)$ is compatible with the meson $\chi_{c1}(2^3P_1)$, with $J^{PC} = 1^{++}$.

\section{Formalism}
\subsection{Self energies}
The Hamiltonian we consider,
\begin{equation}
	\label{eqn:Htot}
	H = H_0 + V  \mbox{ },
\end{equation}
is the sum of an "unperturbed" part, $H_0$, acting only in the bare meson space, and of a second part, $V$, which can couple a meson state to a continuum made up of meson-meson intermediate states. 

The dispersive equation, resulting from a nonrelativistic Schr\"odinger equation, can be written as
\begin{subequations}
\begin{equation}
	\label{eqn:self-a}
	\Sigma(E_a) = \sum_{BC} \int_0^{\infty} q^2 dq \mbox{ } \frac{\left| V_{a,bc}(q) \right|^2}{E_a - E_{bc}}  \mbox{ },
\end{equation}
where the bare energy $E_a$ satisfies: 
\begin{equation}
	\label{eqn:self-trascendental}
	M_a = E_a + \Sigma(E_a)  \mbox{ }.
\end{equation}
\end{subequations}
$M_a$ in Eq. (\ref{eqn:self-trascendental}) is the physical mass of the meson $A$, with self energy $\Sigma(E_a)$. 
In Eq. (\ref{eqn:self-a}) one has to take the contributions from various channels $BC$'s into account. A channel $BC$ is a meson-meson intermediate state, with relative momentum $q$ and quantum numbers $J_{bc}$ and $\ell$ coupled to the total angular momentum of the meson $A$. The matrix element $V_{a,bc}$ of Eq. (\ref{eqn:self-a}) results from the coupling, due to the operator $V$, between the intermediate state $BC$ and the unperturbed quark-antiquark wave function of the meson $A$; $E_{bc} = E_b + E_c$ is the total energy of the channel $BC$, calculated in the rest frame of $A$.
Finally, if the bare energy of the meson $A$, i.e. $E_a$, is greater than the threshold $E_{bc}$, the self energy of Eq. (\ref{eqn:self-a}) contains poles, and is a complex number [see Eq. (\ref{eqn:virtual-Sigma})].

Since the physics of the dynamics depends on the matrix elements $V_{a,bc}(q)$, one has to choose a precise form for the transition operator, $V$, which is responsible for the creation of $q \bar q$ pairs: our choice is that of the unquenched quark model of Ref. \cite{bottomonium}, so a $^{3}P_0$ model.

\subsection{Unquenched quark model}
In the unquenched quark model \cite{bottomonium,Santopinto:2010zza} the effects of $q \bar q$ sea pairs are introduced explicitly into the quark model (QM) through a QCD-inspired $^{3}P_0$ pair-creation mechanism. 
This approach, which is a generalization of the unitarized quark model by T\"ornqvist and Zenczykowski \cite{Tornqvist}, was motivated by later work by Isgur and coworkers on the flux-tube breaking model. They showed that the QM emerges as the adiabatic limit of the flux-tube model to which the effects of $q \bar{q}$ pair creation can be added as a perturbation \cite{Geiger:1996re}. 
Therefore, our approach is based on a QM to which the quark-antiquark pairs with vacuum quantum numbers are added perturbatively. The pair-creation mechanism is inserted at the quark level and the one-loop diagrams are computed by summing over the possible intermediate states. 

Under these assumptions, the meson wave function is made up of a zeroth order quark-antiquark configuration plus a sum over all the possible higher Fock components due to the creation of $^{3}P_0$ quark-antiquark pairs. To leading order in pair creation, the meson wave function is given by 
\begin{eqnarray} 
	\label{eqn:Psi-A}
	\mid \psi_A \rangle &=& {\cal N} \left[ \mid A \rangle 
	+ \sum_{BC \ell J} \int d \vec{q} \, \mid BC \vec{q} \, \ell J \rangle \right.
	\nonumber\\
	&& \hspace{2cm} \left.  \frac{ \langle BC \vec{q} \, \ell J \mid T^{\dagger} \mid A \rangle } 
	{E_a - E_b - E_c} \right] ~, 
\end{eqnarray}
where $T^{\dagger}$ represents the $^{3}P_0$ quark-antiquark pair creation operator \cite{Roberts:1992}, $A$ is the meson, $B$ and $C$ are the intermediate virtual mesons, and $E_a$, $E_b = \sqrt{M_b^2 + q^2}$ and $E_c = \sqrt{M_c^2 + q^2}$ are their respective energies, $\vec{q}$ and $\ell$ the relative radial momentum and orbital angular momentum of $B$ and $C$, and $J$ is the total angular momentum, with $\vec{J} = \vec{J}_b + \vec{J}_c + \vec{\ell}$. 

The $^{3}P_0$ quark-antiquark pair-creation operator of Eq. (\ref{eqn:Psi-A}) can be written as \cite{Roberts:1992}
\begin{equation}
	\label{eqn:Tdag}
	\begin{array}{rcl}
	T^{\dagger} &=& -3 \, \gamma_0 \, \int d \vec{p}_3 \, d \vec{p}_4 \, 
	\delta(\vec{p}_3 + \vec{p}_4) \, C_{34} \, F_{34} \,  
	{e}^{-r_q^2 (\vec{p}_3 - \vec{p}_4)^2/6 }\,  \\
	& & \left[ \chi_{34} \, \times \, {\cal Y}_{1}(\vec{p}_3 - \vec{p}_4) \right]^{(0)}_0 \, 
	b_3^{\dagger}(\vec{p}_3) \, d_4^{\dagger}(\vec{p}_4) ~,   
	\end{array}
\end{equation}
where $b_3^{\dagger}(\vec{p}_3)$ and $d_4^{\dagger}(\vec{p}_4)$ are the creation operators for a quark and an antiquark with momenta $\vec{p}_3$ and $\vec{p}_4$, respectively. The quark and antiquark pair is characterized by a color singlet wave function $C_{34}$, a flavor singlet wave function $F_{34}$, a spin triplet wave function $\chi_{34}$ with spin $S=1$ and a solid spherical harmonic ${\cal Y}_{1}(\vec{p}_3 - \vec{p}_4)$ that indicates that the quark and antiquark are in a relative $P$ wave. 
Since the operator $T^{\dagger}$ creates a pair of constituent quarks with an actual size, the pair creation point has to be smeared out by a gaussian factor, whose width $r_q$ was determined from meson decays to be in the range $0.25-0.35$ fm \cite{Geiger:1996re,Geiger-Isgur,SilvestreBrac:1991pw}. 

The pair-creation strength $\gamma_0$ is a dimensionless constant, fitted to the strong decay widths of $c \bar c$ states (see Sec. \ref{Strong decay widths} for details). 
The matrix elements of the pair-creation operator $T^{\dagger}$ were derived in explicit form in the harmonic oscillator basis as in Ref. \cite{Roberts:1992}, using standard Jacobi coordinates. 
The meson wave functions have good flavor symmetry and depend on a single oscillator parameter $\alpha$, which, according to the literature \cite{Barnes:2007xu,Ackleh:1996yt,Barnes:2005pb}, is taken to be $\alpha=0.50$ GeV.
%%%%%%%%%%%%%%%%%%%%%%%%%%%%%%%%%%%%%%%
\begin{figure}[htbp]
\begin{center}
\includegraphics[width=8cm]{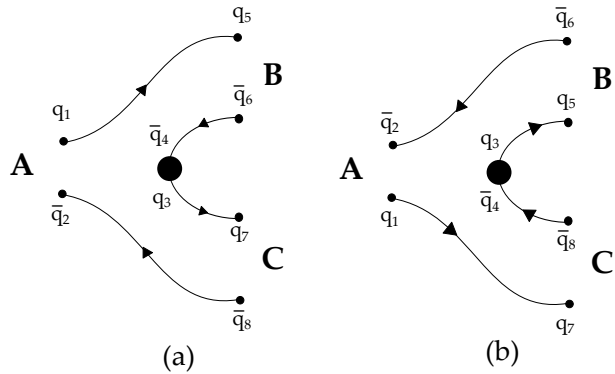}
\end{center}
\caption{Two diagrams can contribute to the process $A\rightarrow BC$. $q_{i}$ and $\bar{q}_{i}$ stand for the various initial ($i=1-4$) and final ($i=5-8$) quarks or antiquarks, respectively. Picture from Ref. \cite{bottomonium}. APS copyright.} 
\label{fig:diagrammi3P0}
\end{figure} 
%%%%%%%%%%%%%%%%%%%%%%%%%%%%%%%%%%%%%%%

In the UCQM, the coupling $V_{a,bc}$ between the continuum channel $BC$ and the unperturbed wave function of the meson $A$ can be calculated as 
\begin{equation}
	V_{a,bc}(q) = \sum_{\ell J} \left\langle BC \vec q  \, \ell J \right| T^\dag \left| A \right\rangle  \mbox{ }.
\end{equation}
In general, two different diagrams can contribute to the transition matrix element $\left\langle BC \vec q  \, \ell J \right| T^\dag \left| A \right\rangle $ (see Fig. \ref{fig:diagrammi3P0}): in the first one, the quark in $A$ ends up in $B$, while in the second one it ends up in $C$. In the majority of cases, one of these two diagrams vanishes; however, for some matrix elements, both must be taken into account \cite{bottomonium}, as for example, this is the case of the coupling $\eta_c \rightarrow J/\Psi J/\Psi$, where the initial $\left| c \bar c \right\rangle$ state is coupled to the final state $\left| c \bar c; c \bar c \right\rangle$ and the created pair is a $c \bar c$ one.   

Finally, the expression for the self energy of the meson $A$, Eq. (\ref{eqn:self-a}), can be re-written as 
\begin{equation}
	\label{eqn:Sigma_A}
	\Sigma(E_a) = \sum_{BC \ell J} \int_0^{\infty} q^2 dq \mbox{ } 
	\frac{\left|\left\langle BC \vec q  \, \ell J \right| T^\dag \left| A \right\rangle \right|^2} {E_a - E_b - E_c} 
	\mbox{ }.
\end{equation}

\subsection{Godfrey and Isgur's relativized quark model}
\label{Godfrey and Isgur's relativized constituent quark model}
There is a huge number of studies on meson spectroscopy, based on different pictures for mesons; they include $q \bar q$ mesons \cite{Godfrey:1985xj,Iachello:1991re,Eichten:1974af}, meson-meson molecules \cite{Danilkin:2010cc,Hanhart:2007yq,Baru:2011rs,Fulsom:2007ve,Weinstein:1990gu,Barnes:1991em,Tornqvist:1993ng}, tetraquarks \cite{Jaffe:1976ig,Maiani:2004uc,Santopinto:2006my} and quarkonium hybrids \cite{LlanesEstrada:2000hj,Buisseret:2006wc,Guo:2008yz} and references can be found in review papers like \cite{REVIEW}.

The relativized QM by Godfrey and Isgur \cite{Godfrey:1985xj} is a potential model for $q \bar q$ meson spectroscopy. 
This model assumes a relativistic dispersion relation for the quark kinetic energy, a QCD-motivated running coupling constant $\alpha_s(r)$, a flavor-dependent potential smearing parameter $\sigma$, and replaces factors of quark mass with quark kinetic energy. 

The Hamiltonian of the model \cite{Godfrey:1985xj} is given by
\begin{equation}
	\label{eqn:GI-Hamiltonian}
	H = \sqrt{q^2 + m_1^2} + \sqrt{q^2 + m_2^2} + V_{\mbox{conf}} + V_{\mbox{hyp}} + V_{\mbox{so}}  \mbox{ },
\end{equation}
where $m_1$ and $m_2$ are the masses of the constituent quark and antiquark inside the meson, $q$ is their relative momentum (with conjugate coordinate $r$), $V_{\mbox{conf}}$, $V_{\mbox{hyp}}$ and $V_{\mbox{so}}$ are the confining, hyperfine and spin-orbit potentials, respectively.

The confining potential,
\begin{equation}
	V_{\mbox{conf}} = - \left(\frac{3}{4} \mbox{ } c + \frac{3}{4} \mbox{ } br -\frac{\alpha_s(r)}{r} \right) 
	\vec F_1 \cdot \vec F_2  \mbox{ },
\end{equation}
contains a constant, $c$, a linear confining term and a Coulomb-like interaction, depending on the renormalized running coupling constant of QCD, $\alpha_s(r)$ (for more details see Ref. \cite{Godfrey:1985xj}); moreover, one has:
\begin{equation}
	\left\langle q \bar q \right| \vec F_1 \cdot \vec F_2 \left| q \bar q \right\rangle = - \frac{4}{3}  \mbox{ }.
\end{equation}

The hyperfine interaction is written as \cite{Godfrey:1985xj}
\begin{equation}
	\label{eqn:Vhyp}
	\begin{array}{rcl}
	V_{\mbox{hyp}} & = & -\frac{\alpha_s(r)}{m_{1}m_{2}} \left[\frac{8\pi}{3} \vec S_{1} \cdot \vec S_{2} \mbox{ }
	\delta ^{3}(\vec r) \right. \\ 
	& + & \left. \frac{1}{r^{3}} \left( \frac{3 \mbox{ } \vec S_{1} \cdot \vec r \mbox{ } 
	\vec S_{2} \cdot \vec r}{r^{2}} - \vec S_{1} \cdot \vec S_{2}\right) \right] \mbox{ } 
	\vec F_{i} \cdot \vec F_{j}  \mbox{ }.
	\end{array}
\end{equation}

The spin-orbit potential \cite{Godfrey:1985xj}, 
\begin{equation}
	V_{\mbox{so}} = V_{\mbox{so,cm}} + V_{\mbox{so,tp}}  \mbox{ },
\end{equation}
is the sum of two contributions, where
\begin{subequations}
\begin{equation}
	\begin{array}{rcl}
	V_{\mbox{so,cm}} & = & - \frac{\alpha_s(r)}{r^{3}} \left( \frac{1}{m_{i}}+\frac{1}{m_{j}} \right) \\
	& & \left( \frac{\vec S_{i}}{m_{i}}+\frac{\vec S_{j}}{m_{j}} \right) \cdot \vec L 
	\;\;\vec F_{i}\cdot \vec F_{j}
	\end{array}
\end{equation}
is the color-magnetic term and
\begin{equation}
	V_{\mbox{so,tp}} = - \frac{1}{2r}\frac{\partial H_{ij}^{conf}}{\partial r} \left( \frac{\vec S_{i}} 
	{m_{i}^{2}}+\frac{\vec S_{j}}{m_{j}^{2}}\right) \cdot \vec L
\end{equation}
\end{subequations}
is the Thomas-precession term. 

\section{Results}
\subsection{Strong decay widths}
\label{Strong decay widths} 
In this section, we show our results for the strong decay widths of $3S$, $2P$, $1D$ and $2D$ charmonium states above the $D \bar D$ threshold (see Table \ref{tab:strong-decays}). 

The decay widths are calculated within the $^3P_0$ model \cite{SilvestreBrac:1991pw,Ackleh:1996yt,Barnes:2005pb} as
\begin{equation}
	\Gamma_{A \rightarrow BC} = \Phi_{A \rightarrow BC}(q_0) \sum_{\ell, J} 
	\left| \left\langle BC \vec q_0  \, \ell J \right| T^\dag \left| A \right\rangle \right|^2 \mbox{ }.
\end{equation}
$\Phi_{A \rightarrow BC}(q_0)$ is the standard relativistic phase space factor \cite{Ackleh:1996yt,Barnes:2005pb}, 
\begin{equation}
	\label{eqn:relPSF}
	\Phi_{A \rightarrow BC} = 2 \pi q_0 \frac{E_b(q_0) E_c(q_0)}{M_a}  \mbox{ },
\end{equation}
depending on the relative momentum $q_0$ between $B$ and $C$ and on the energies of the two intermediate state mesons, $E_b = \sqrt{M_b^2 + q_0^2}$ and $E_c = \sqrt{M_c^2 + q_0^2}$ (for the values of $M_b$ and $M_c$, see Table \ref{Masses-non-cc}).
%%%%%%%%%%%%%%%%%%%%%%%%%%%%%%%%%%%%%%%%%%%%%%%%%%%%%%%%%%%%%%%%%%%%%
\begin{table}[htbp]  
\begin{center}
\begin{tabular}{ccc} 
\hline 
\hline \\
State               & Mass [GeV] & Source                  \\ \\ \hline \\
$D$                 &    1.867   & \cite{Nakamura:2010zzi} \\
$D^*(2007)$         &    2.009   & \cite{Nakamura:2010zzi} \\
$D_s$               &    1.969   & \cite{Nakamura:2010zzi} \\
$D_s^*$             &    2.112   & \cite{Nakamura:2010zzi} \\   \\
\hline 
\hline
\end{tabular}
\end{center}
\caption{Masses of open charm mesons used in the calculations.} 
\label{Masses-non-cc}  
\end{table}
%%%%%%%%%%%%%%%%%%%%%%%%%%%%%%%%%%%%%%%%%%%%%%%%%%%%%%%%%%%%%%%%%%%%%
The operator $T^\dag$ inside the $^{3}P_0$ amplitudes $\left\langle BC	\vec q_0  \, \ell J \right| T^\dag \left| A \right\rangle$ is that of Eq. (\ref{eqn:Tdag}), which also contains the quark form factor of Refs. \cite{Geiger:1996re,Geiger-Isgur}. 
The introduction of this quark form factor, which is just a Gaussian function in the relative momentum between the quark and the antiquark of the created pair, in the $^{3}P_0$ model transition operator determines slightly different values for the model parameters (see Table \ref{tab:parameters}). 
Specifically, the value of the pair-creation strength $\gamma_0$, which is fitted to the reproduction of the experimental strong decay widths of Table \ref{tab:strong-decays}, is greater than that which would be obtained in the standard $^{3}P_0$ model \cite{Ackleh:1996yt,Barnes:2005pb}, i.e. $\gamma_0 = 0.4$. 
%%%%%%%%%%%%%%%%%%%%%%%%%%%%%%%%%%%%%%%%%%%%%%%%%%%%%%%%%%%%%%%%%%%%%%%%%% 
\begin{table}[h]  
\begin{center}
\begin{tabular}{cc} 
\hline 
\hline \\
Parameter  &  Value     \\ \\
\hline \\
$\gamma_0$ & 0.510       \\  
$\alpha$   & 0.500 GeV   \\  
$r_q$      & 0.335 fm    \\
$m_n$      & 0.330 GeV   \\
$m_s$      & 0.550 GeV   \\
$m_c$      & 1.50 GeV    \\   \\
\hline 
\hline
\end{tabular}
\end{center}
\caption{Parameters of $^3P_0$ the model.}
\label{tab:parameters}  
\end{table}

\begin{table*}
\begin{tabular}{ccccccccc} 
\hline 
\hline \\
State                & $DD$ & $DD^*$ & $D^*D^*$ & $D_sD_s$ & $D_sD_s^*$ & $D_s^*D_s^*$ & Total & Exp. \\ \\
\hline \\
$\eta_c(3^1S_0)$     &  --  & 38.8 & 52.3 &  --  &  --  &  --  & 91.1  &  --  \\  
$\Psi(4040)(3^3S_1)$ &  0.2 & 37.2 & 39.6 & 3.3  &  --  &  --  & 80.3  & $80\pm10$ \\  
$h_c(2^1P_1)$        &  --  & 64.6 &  --  &  --  &  --  &  --  & 64.6  &  --  \\  
$\chi_{c0}(2^3P_0)$  & 97.7 &  --  &  --  &  --  &  --  &  --  & 97.7  &  --  \\  
$\chi_{c2}(2^3P_2)$  & 27.2 &  9.8 &  --  &  --  &  --  &  --  & 37.0  &  --  \\  
$\Psi(3770)(1^3D_1)$ & 27.7 &  --  &  --  &  --  &  --  &  --  & 27.7  & $27.2\pm1.0$ \\   
$c \bar c(1^3D_3)$   & 1.7  &  --  &  --  &  --  &  --  &  --  & 1.7   &  --  \\  
$c \bar c(2^1D_2)$   &  --  & 62.7 & 46.4 &  --  &  8.8 &  --  & 117.9 &  --  \\  
$\Psi(4160)(2^3D_1)$ & 11.2 &  0.4 & 39.4 &  2.1 &  5.6 &  --  & 58.7  & $103\pm8$ \\  
$c \bar c(2^3D_2)$   &  --  & 43.5 & 49.3 &  --  & 11.3 &  --  & 104.1 &  --  \\  
$c \bar c(2^3D_3)$   & 17.2 & 58.3 & 48.1 &  3.6 &  2.6 &  --  & 129.8 &  --  \\ \\
\hline 
\hline
\end{tabular}
\caption{Strong decay widths (in MeV) for $3S$, $2P$, $1D$ and $2D$ charmonium states. The values of the model parameters are given in Table \ref{tab:parameters}. The symbol -- in the table means that a certain decay is forbidden by selection rules or that the decay cannot take place because it is below threshold.} 
\label{tab:strong-decays}  
\end{table*}
%%%%%%%%%%%%%%%%%%%%%%%%%%%%%%%%%%%%%%%%%%%%%%%%%%%%%%%%%%%%%%%%%%%%%

Another difference between our calculation and those of Refs. \cite{Ackleh:1996yt,Barnes:2005pb} is the substitution of the pair-creation strength $\gamma_0$ with the effective strength $\gamma_0^{\mbox{eff}}$ of App. \ref{Effective strength gamma0-eff}. 
The introduction of this effective mechanism suppresses those diagrams in which a heavy $q \bar q$ pair is created. 
More details on this mechanism can be found in Refs. \cite{bottomonium,Kalashnikova:2005ui}.

Finally, the results of our calculation, obtained with the values of the model parameters of Table \ref{tab:parameters}, are reported in Table \ref{tab:strong-decays}. 
This set of parameters is also used in the self energy calculation of Sec. \ref{Bare energy calculation within the relativized constituent quark model} in order to compute the vertices $\left\langle BC \vec q  \, \ell J \right| T^\dag \left| A \right\rangle$ of Eqs. (\ref{eqn:Sigma_A}) and (\ref{eqn:virtual-Sigma}).

\subsection{Bare energy calculation within the relativized quark model. Self energies of $c \bar c$ states}
\label{Bare energy calculation within the relativized constituent quark model}
The relativized QM \cite{Godfrey:1985xj}, which is described in Sec. \ref{Godfrey and Isgur's relativized constituent quark model}, is here used to compute the bare energies of the $c \bar c$ states that we need in the self energy calculation. 
%%%%%%%%%%%%%%%%%%%%%%%%%%%%%%%%%%%%%%%%%%%%%%%%%%%%%%%%%%%%%%%%%%%%%
\begin{table}[htbp]  
\begin{center}
\begin{tabular}{llllll}
\hline
\hline \\
$m_c$                    & = 1.562 GeV & $b$                & = 0.1477 GeV$^2$ & $\alpha_s^{\mathrm{cr}}$ & = 0.600    \\
$\Lambda$            & = 0.200 GeV & $c$                 & = 0.069 GeV           & $\sigma_0$                      & = 1.463 GeV \\ 
$s$                        & = 2.437        & $\epsilon_c$   & = $-$0.2500          & $\epsilon_t$                     & = 0.0300   \\ 
$\epsilon_{so(V)}$ & = $-$0.0314 & $\epsilon_{so(S)}$ & = 0.0637         &                          &            \\ \\
\hline
\hline
\end{tabular}
\end{center}
\caption{Values of Godfrey and Isgur's model parameters, obtained by fitting the results of Eq. (\ref{eqn:self-trascendental}) to the experimental data \cite{Nakamura:2010zzi}.}
\label{tab:Parametri-self}
\end{table}
%%%%%%%%%%%%%%%%%%%%%%%%%%%%%%%%%%%%%%%%%%%%%%%%%%%%%%%%%%%%%%%%%%%%%
In our study, we computed the bare energies $E_a$'s of Eq. (\ref{eqn:self-trascendental}) as the eigenvalues of Eq. (\ref{eqn:GI-Hamiltonian}).
At variance with QM calculations, such as that of Ref. \cite{Godfrey:1985xj}, we did not fit the eigenvalues of Eq. (\ref{eqn:GI-Hamiltonian}) to the experimental data \cite{Nakamura:2010zzi}.
In our case, the quantities fitted to the spectrum of charmonia \cite{Nakamura:2010zzi} are the masses $M_a$'s of Eq. (\ref{eqn:self-trascendental}) and therefore the fitting procedure is an iterative one. 
Our resulting values for the parameters of Godfrey and Isgur's model are shown in Table \ref{tab:Parametri-self}.

Once the values of the bare energies are known, it is possible to calculate the self energies $\Sigma(E_a)$'s of $1S$, $2S$, $1P$, $2P$ and $1D$ $c \bar c$ states through Eq. (\ref{eqn:Sigma_A}). 
If the bare energy of the meson $A$ is above the threshold $BC$, i.e. $E_a > M_b + M_c$, the contribution to the self energy due to the meson-meson channel $BC$ is computed as
\begin{equation}
	\label{eqn:virtual-Sigma}
	\begin{array}{l}
	\Sigma(E_a;BC) \\
	\hspace{0.5cm} = \mbox{ } \mathcal{P} \int_{M_b+M_c}^{\infty} \frac{dE_{bc}}{E_a - E_{bc}} \mbox{ } 
	\frac{q E_b E_c}{E_{bc}} \left| \left\langle BC \vec q \, \ell J \right| T^\dag \left| A \right\rangle \right|^2 \\
	\hspace{0.5cm} + \mbox{ } 2 \pi i \left\{ \frac{q E_b E_c}{E_a} 
	\left| \left\langle BC \vec q \, \ell J \right| T^\dag \left| A \right\rangle \right|^2 \right\}_{E_{bc} = E_a} 
	\mbox{ },
	\end{array}
\end{equation}
where the symbol $\mathcal{P}$ represents the principal part integral, which can be computed numerically, and $2 \pi i \left\{ \frac{q E_b E_c}{E_a} \left| \left\langle BC \vec q \, \ell J \right| T^\dag \left| A \right\rangle \right|^2 \right\}_{E_{bc} = E_a}$ is the imaginary part of the self energy, related to the decay width by:
\begin{equation}
	\Gamma_{A \rightarrow BC} = \mbox{Im}\left[\Sigma(E_a;BC)\right]  \mbox{ }.
\end{equation}	

\begin{figure}[htbp]
\begin{center}
\includegraphics[width=8cm]{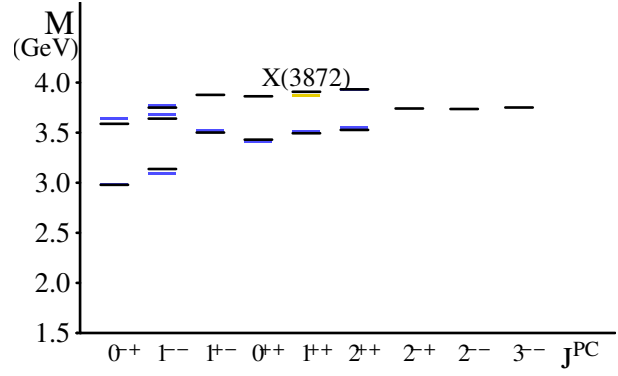}
\end{center}
\caption{Comparison between the calculated masses (black lines) of $1S$, $2S$, $1P$, $2P$ and $1D$ charmonium states via Eq. (\ref{eqn:self-trascendental}) and the experimental ones \cite{Nakamura:2010zzi} (blue boxes). The experimental mass of the $X(3872)$ is indicated by a yellow box. The new values of the parameters of Godfrey and Isgur's model are taken from Table \ref{tab:Parametri-self}.} 
\label{fig:charmonium-spectrum-self}
\end{figure}

Finally, the results of our UCQM calculation, obtained with the set of parameters of Tables \ref{tab:parameters} and \ref{tab:Parametri-self} and with the effective pair-creation strength $\gamma_0^{\mbox{eff}}$ of App. \ref{Effective strength gamma0-eff}, are shown in Table \ref{tab:Mass-shifts} and Fig. \ref{fig:charmonium-spectrum-self}. 
%%%%%%%%%%%%%%%%%%%%%%%%%%%%%%%%%%%%%%%%%%%%%%%%%%%%%%%%%%%%%%%%%%%%%

\begin{table*}
\begin{tabular}{cccccccccccccccc} 
\hline 
\hline \\
State                & $J^{PC}$ & $D\bar D$ & $\bar DD^*$  & $\bar D^*D^*$ & $D_s \bar D_s$ & $D_s \bar D_s^*$ & $D_s^* \bar D_s^*$ & $\eta_c\eta_c$ & $\eta_cJ/\Psi$ & $J/\Psi J/\Psi$ & $\Sigma(E_a)$ & $E_a$ & $M_a$ & $M_{exp.}$  \\
                     &          &           & $D \bar D^*$ &             &                  & $\bar D_s D_s^*$ &                    &                &              &                   &       &         &  &  & \\ \\
\hline \\
$\eta_c(1^1S_0)$     & $0^{-+}$ & --  & -34   & -31  & --  & -8  & -8  & --  & --  & -2  & -83  & 3062 & 2979 & 2980 \\
$J/\Psi(1^3S_1)$     & $1^{--}$ & -8  & -27   & -41  & -2  & -6  & -10 & --  & -2  & --  & -96  & 3233 & 3137 & 3097 \\
$\eta_c(2^1S_0)$     & $0^{-+}$ & --  & -52   & -41  & --  & -9 & -8  & --  & --  & -1  & -111 & 3699 & 3588 & 3637 \\
$\Psi(2^3S_1)$       & $1^{--}$ & -18 & -42   & -54  & -2  & -7  & -10 & --  & -1  & --  & -134 & 3774 & 3640 & 3686 \\
$h_c(1^1P_1)$        & $1^{+-}$ & --  & -59   & -48  & --  & -11 & -10 & --  & -2  & --  & -130 & 3631 & 3501 & 3525 \\
$\chi_{c0}(1^3P_0)$  & $0^{++}$ & -31 &  --   & -72  & -4  &  -- & -15 &  0  & --  & -3  & -125 & 3555 & 3430 & 3415 \\ 
$\chi_{c1}(1^3P_1)$  & $1^{++}$ & --  &  -54  & -53  & --  & -9  & -11 & --  & --  & -2  & -129 & 3623 & 3494 & 3511 \\
$\chi_{c2}(1^3P_2)$  & $2^{++}$ & -17 &  -40  & -57  & -3  & -8  & -10 &  0  & --  & -2  & -137 & 3664 & 3527 & 3556 \\
$h_c(2^1P_1)$        & $1^{+-}$ & --  & -55   & -76  & --  & -12 & -8  & --  & -1  & --  & -152 & 4029 & 3877 & --  \\
$\chi_{c0}(2^3P_0)$  & $0^{++}$ & -23 &  --   & -86 & -1   &  -- & -13 &  0  & --  & -1  & -124 & 3987 & 3863 & -- \\ 
$\chi_{c1}(2^3P_1)$  & $1^{++}$ & --  &  -30  & -66  & --  & -11 & -9 & --  & --  & -1  & -117 & 4025 & 3908 & 3872 \\
$\chi_{c2}(2^3P_2)$  & $2^{++}$ & -2  &  -42  & -54  & -4  &  -8 & -10 &  0  & --  & -1  & -121 & 4053 & 3932 & 3927 \\
$c \bar c(1^1D_2)$   & $2^{-+}$ & --  &  -99  & -62 & --  & -12 & -10 & --  & --  & -1  & -184 & 3925 & 3741 & -- \\    
$\Psi(3770)(1^3D_1)$ & $1^{--}$ & -11 &  -40  & -84  & -4  &  -2 & -16 & --  &  0  & --  & -157 & 3907 & 3750 & 3775 \\   
$c \bar c(1^3D_2)$   & $2^{--}$ & --  &  -106 & -61  & --  & -11 & -11 & --  & -1  & --  & -190 & 3926 & 3736 & -- \\     
$c \bar c(1^3D_3)$   & $3^{--}$ & -25 &  -49  & -88 & -4  & -8 & -10 & --  & -1  & --  & -185 & 3936 & 3751 & -- \\ \\
\hline 
\hline
\end{tabular}
\caption{Self energies, $\Sigma(E_a)$ (in MeV, see column 12), for charmonium states due to coupling to the meson-meson continuum, calculated with the effective pair-creation strength of Eq. (\ref{eqn:gamma0-eff}) and the values of the UCQM parameters of Table \ref{tab:parameters}. Columns 3-11 show the contributions to $\Sigma(E_a)$ from various channels $BC$, such as $D \bar D$, $D \bar D^*$ and so on. In column 13 are reported the values of the bare energies, $E_a$, calculated within the relativized QM \cite{Godfrey:1985xj}, with the values of the model parameters of Table \ref{tab:Parametri-self}. 
In column 14 are reported the theoretical estimations $M_a$ of the masses of the $c \bar c$ states, which are the sum of the self energies $\Sigma(E_a)$ and the bare energies $E_a$. 
Finally, in column 15 are reported the experimental values of the masses of the $c \bar c$ states, as from the PDG \cite{Nakamura:2010zzi}.} 
\label{tab:Mass-shifts}  
\end{table*}
%%%%%%%%%%%%%%%%%%%%%%%%%%%%%%%%%%%%%%%%%%%%%%%%%%%%%%%%%%%%%%%%%%%%%

\subsection{Nature of the $X(3872)$ resonance}
The quark structure of the $X(3872)$ resonance, observed for the first time by the Belle Collaboration in the decay of the $B$ meson \cite{Choi:2003ue} and then confirmed by CDF \cite{Acosta:2003zx}, D0 \cite{Abazov:2004kp} and BABAR \cite{Aubert:2004ns}, still remains an open puzzle. 
Indeed, at the moment, there are two possible interpretations for the meson: a weakly bound $1^{++}$ molecule \cite{Danilkin:2010cc,Hanhart:2007yq,Baru:2011rs,Swanson:2003tb,Fulsom:2007ve} or a charmonium state, with $1^{++}$ or $2^{-+}$ quantum numbers \cite{Meng:2007cx}. For sure, we can only say that the decay channels where it was observed imply $1^{++}$ or $2^{-+}$ quantum numbers \cite{Abe:2005ix}, while the other hypotheses are excluded by more than $3 \sigma$ \cite{Abulencia:2005zc}.  
It is thus necessary, in order to study properties of the $X(3872)$ such as the decay modes, to make an assumption regarding its quark structure that is compatible with the quantum numbers $1^{++}$ or $2^{-+}$.

The first and easiest possibility is to consider the $X(3872)$ as a $c \bar c$ state \cite{Meng:2007cx}. 
In this case, the $X(3872)$ would correspond to a $2^3P_1$ resonance [$\chi_{c1}(2P)$, $J^{PC}=1^{++}$] or to a $1^1D_2$ ($J^{PC}=2^{-+}$) one, according to the estimations of the QM \cite{Godfrey:1985xj,Barnes:2005pb}. 
Indeed, QM predictions show that $2^3P_1$ and $1^1D_2$ states are the only ones compatible with $1^{++}$ or $2^{-+}$ quantum numbers and lying approximately in the same energy region as the $X(3872)$.
The relativized QM \cite{Godfrey:1985xj} predicts these states to be at energies of 3.95 and 3.84 GeV, respectively.
However, the most recent results by the LHCb collaboration \cite{Aaij:2013zoa} seem to favor the $1^{++}$ quantum numbers.

Our idea is thus to see whether the introduction of loop corrections into the QM can help clarify the problem of the nature of the $X(3872)$.
Indeed, we think that the uncommon properties of the $X(3872)$ are due to its proximity to the $D \bar D^*$ decay threshold and cannot easily be explained within a standard quark-antiquark picture for mesons. 

In our calculation of Table \ref{tab:Mass-shifts}, we have re-fitted the spectrum of charmonia through Eq. (\ref{eqn:self-trascendental}); here, the mass of a meson results from the sum of a bare energy term computed within the relativized QM of Ref. \cite{Godfrey:1985xj}, with a self energy correction computed within the unquenched quark model formalism of Refs. \cite{bottomonium,Santopinto:2010zza}.
According to our results for the masses of the $2^3P_1$ and $1^1D_2$ states, i.e. 3.908 and 3.741 GeV, respectively, the $X(3872)$ is compatible with the meson $\chi_{c1}(2P)$ and includes an extra component due to the coupling to the meson-meson continuum, which is responsible for the downward energy shift. 

The second possibility is to treat the $X(3872)$ as a $D \bar D^*$ molecular state with $1^{++}$ quantum numbers \cite{Hanhart:2007yq,Danilkin:2010cc}. 
According to Refs. \cite{Voloshin:1976ap}, the $D \bar D^*$ system with $1^{++}$ quantum numbers can be found by pion exchange and forms a meson molecule. 
More recent molecular model calculations \cite{Tornqvist:2004qy}, including quark exchange kernels for the transitions $D \bar D^* \rightarrow \rho J/\Psi$, $\omega J/\Psi$ in order to predict the $\omega J/\Psi$ decay mode of the $X(3872)$ \cite{Swanson:2003tb}, predict large isospin mixing due to the mass difference between $D^0 \bar D^{*0}$ and $D^+ \bar D^{*-}$.
Nevertheless, in Ref. \cite{Hanhart:2007yq} the authors observe that the one-pion exchange binding mechanism should be taken with greater caution in the $D \bar D^*$ case than in the $NN$ case (see also Refs. \cite{Suzuki:2005ha,Braaten:2007ct,Fleming:2007rp}). 

Another important test for the properties of the $X(3872)$ consists of estimating its strong and radiative decay rates \cite{Meng:2007cx,Suzuki:2005ha,Swanson:2003tb}. 
In Ref. \cite{Meng:2007cx}, the authors re-examine the re-scattering mechanism for the X(3872), which decays to $J/\psi  \rho(\omega)$ through the exchange of $D(^*)$ mesons between intermediate states $D(\bar D)$ and $\bar D^* (D^*)$.
Their results for the ratio $R_{\rho/\omega} \approx 1$, between the decay modes $X(3872) \rightarrow J/\psi  \rho$ and $X(3872) \rightarrow J/\psi \omega$, and for the rate $X(3872) \rightarrow D^0 \bar D^0 \pi^0$, favor a charmonium $c \bar c$ interpretation for the $X(3872)$. 
In Ref. \cite{Suzuki:2005ha}, the author uses semi-quantitative methods to study some properties of the $X(3872)$; he points out that the binding mechanism and the production rates are incompatible with the molecule interpretation.
This is also suggested by the CDF II paper \cite{Bauer:2004bc} where the authors observe  also prompt production and  discuss that a meson-meson molecule with a dimension of a few fm and intrinsic fragility cannot be prompt produced.
By contrast, Refs. \cite{Swanson:2003tb,Danilkin:2010cc,Hanhart:2007yq,Baru:2011rs} suggest a molecular interpretation for the $X(3872)$.

Finally, we do not think that our arguments can, on their own, clarify the picture of the $X(3872)$ resonance completely. 
Thus, it will be necessary to analyze other properties of this meson, such as strong and electromagnetic decays, in order to draw a definitive conclusion. 
In particular, we intend to calculate some of these observables within the UCQM \cite{bottomonium,Santopinto:2010zza}, by also taking the contribution of $q \bar q$ sea pairs into account \cite{FSV4}.

\subsection{Discussion of the results}
In this paper we have presented the results of an unquenched quark model calculation of the self energy corrections to the spectrum of $1S$, $2S$, $1P$, $2P$ and $1D$ charmonium states. In the unquenched quark model, developed in the baryon sector in Refs. \cite{Santopinto:2010zza} and in the meson sector in Refs. \cite{bottomonium}, the effects of quark-antiquark sea pairs are introduced explicitly into the QM through a QCD-inspired $^3P_0$ pair-creation mechanism. 
The UCQM model parameters are fitted to the reproduction of strong decay widths, as is shown in Sec. \ref{Strong decay widths}.

The self energies are corrections to the bare meson masses arising from the coupling to the meson-meson continuum. 
Neglected in naive QM's, these loop effects provide an estimation of the quality of the quenched approximation used in QM calculations in which only valence quarks are taken into account. 
Something similar also happens in the case of lattice QCD, where one has to unquench the calculations in order to evaluate the contribution of the sea quarks to a certain observable.
Therefore, one could say that these kinds of studies can be thought of as tests of the QM and of its range of applicability, and also as an enlargement of the model. 
Several studies on the goodness of the quenched approximation in the QM have already been conducted, such as those of Refs. \cite{Geiger-Isgur,Santopinto:2010zza,bottomonium}, in both the baryon and meson sectors.
If the departure from the QM results is substantial, one can see new physics emerging or better extra degrees of freedom. 
This is the case of the $X(3872)$, which in our picture can be described as a $c \bar c$ state plus higher Fock components mainly due to $D \bar D^*$ and $D^* \bar D^*$ loops.

Our results for the self energies of charmonia show that the pair-creation effects on the spectrum of heavy mesons are relatively small. 
Specifically for charmonium states, they are of the order of $2 - 6\%$, while we have shown in Refs. \cite{bottomonium} that the bottomonium mass shifts induced by the coupling to the meson-meson continuum are less than approximately 1\%. 
The relative mass shifts, i.e. the difference between the self energies of two meson states, are in the order of a few tens of MeV. 
However, as QM's can predict the meson masses with relatively high precision in the heavy quark sector -- higher than can be obtained in the light meson sector or in baryon spectroscopy -- even these corrections can become significant, such as in the case of the $X(3872)$. 

It is interesting that the relative contribution of these corrections to meson masses decreases as the masses of the constituent quarks involved in the calculation increase. Moreover, $q \bar q$ pair creation is a relativistic effect, i.e. more important for low energy states.
This is why we think that it would be quite interesting to use this formalism in the study of light mesons, for which relativistic effects, including $q \bar q$ pair creation, could make important corrections to the meson masses.

\begin{acknowledgments}
This work was supported in part by INFN and in part by Fondazione Angelo Della Riccia, Firenze, Italy and CONACYT (Grant No. 78833) and by PAPIIT-DGAPA (Grant No. IN113711), Mexico.
\end{acknowledgments}

\begin{appendix}

\section{SU$_{\mbox{f}}$(4) couplings}
The SU$_{\mbox{f}}$(4) flavor couplings that we have to calculate in the $^3P_0$ model are  $\langle F _{B}(14)F _{C}(32)|F _{A}(12)F _{0}(34)\rangle$ for the first diagram of Fig. \ref{fig:diagrammi3P0}, and $\langle F _{B}(32)F _{C}(14)|F _{A}(12)F _{0}(34)\rangle$ for the second diagram, where $F_{X}(ij)$ represents the flavor wave function for the meson $X$ (i.e. the initial meson $A$, the final mesons $B$ and $C$ or the $^3P_0$ created pair $0$) made up of the quarks $i$ and $j$.
These overlaps can be easily calculated if we adopt a matrix representation of the mesons \cite{Ono:1983rd}. 
In this case, the two diagrams become, respectively,
\begin{equation}
	\begin{array}{rcl}
	\langle F _{B}(14)F _{C}(32)|F _{A}(12)F _{0}(34)\rangle & = & Tr[F _{A}F _{B}^{T}F _{0}F _{C}^{T}] \\ 
	& = & \frac{1}{2}Tr[F _{A}F _{B}^{T}F _{C}^{T}] \mbox{ }, \\
	\langle F _{B}(32)F _{C}(14)|F _{A}(12)F _{0}(34)\rangle & = & Tr[F _{A}F _{C}^{T}F _{0}F _{B}^{T}] \\ 
	& = & \frac{1}{2}Tr[F _{A}F _{C}^{T}F _{B}^{T}] \mbox{ }.
	\end{array}
\end{equation}

For the SU$_{\mbox{f}}$(5) flavor couplings, which have already been used for the bottomonium self energies in a preceding paper \cite{bottomonium}, the formulas are
\begin{equation}
	\begin{array}{rcl}
	\langle F _{B}(14)F _{C}(32)|F _{A}(12)F _{0}(34)\rangle & = & Tr[F _{A}F _{B}^{T}F _{0}F _{C}^{T}] \\ 
	& = & \frac{1}{\sqrt{5}}Tr[F _{A}F _{B}^{T}F _{C}^{T}] \mbox{ }, \\
	\langle F _{B}(32)F _{C}(14)|F _{A}(12)F _{0}(34)\rangle & = & Tr[F _{A}F _{C}^{T}F _{0}F _{B}^{T}] \\
	& = & \frac{1}{\sqrt{5}}Tr[F _{A}F _{C}^{T}F _{B}^{T}] \mbox{ }.
	\end{array}
\end{equation}

\section{Effective strength $\gamma_0^{\mbox{eff}}$}
\label{Effective strength gamma0-eff}
It is known that the standard $^3P_0$ model should not be applied for heavy-quark pair creation \cite{bottomonium}; alternatively, the contribution from heavy channels should somehow be suppressed. 
Thus, in order to minimize the contributions from $c \bar c$ loops in Eq. (\ref{eqn:Sigma_A}), we use the modified pair-creation mechanism of Refs. \cite{Kalashnikova:2005ui,bottomonium}. 
This involves substituting the pair-creation strength of the $^3P_0$ model, $\gamma_0$, with an effective strength, $\gamma_0^{\mbox{eff}}$, defined as
\begin{equation}
	\label{eqn:gamma0-eff}
	\gamma_0^{\mbox{eff}} = \frac{m_n}{m_i} \mbox{ } \gamma_0  \mbox{ },
\end{equation}
with $i$ = $n$ (i.e. $u$ or $d$), $s$, $c$ and $b$ (see Table \ref{tab:parameters}). 

\end{appendix}

%%%%%%%%%%%%%%%%%%%%%%%%%%%%%%%%%%%%%%%%%%%%%%%%%%

\end{document}